# Structure of Clifford-Weyl Algebras And

# Representations of ortho-symplectic Lie Superalgebras

Nasser Boroojerdian

Department of Mathematics and Computer Science,

Amirkabir University of Technology, Tehran, Iran

Email:broojerd@aut.ac.ir

**Abstract**

In this article, the structure of the Clifford-Weyl superalgebras and their associated Lie superalgebras will be investigated. These superalgebras have a natural supersymmetric inner product which is invariant under their Lie superalgebra structures. The Clifford-Weyl superalgebras can be realized as tensor product of the algebra of alternating and symmetric tensors respectively, on the even and odd parts of their underlying superspace. For Physical applications in elementary particles, we add star structures to these algebras and investigate the basic relations. Ortho-symplectic Lie algebras are naturally present in these algebras and their representations on these algebras can be described easily.

CONTENTS

## 1. Introduction

In this article, everything starts from a supersymmetric bilinear form on a super vector space. These spaces have an associated super algebra, called their Clifford-Weyl algebra that is realized as the space of tensor product of the algebra of alternating and symmetric tensors on the underlying space. Therefore, we will first give an overview of the basic concepts of the structure of the alternating and symmetric tensor algebras, then after fixing the notations, we will investigate the structure of the Clifford-Weyl algebras.

First, we will find the basic properties of the structure of Clifford's and Weyl's algebras separately, and then we will extend the obtained results in a unified algebra on superspaces. These algebras have a natural inner product that is compatible with their algebras and Lie algebras structures.

**Symmetric and alternating inner products**: All vector spaces discussed in this article are real or complex vector spaces of finite dimension. On a vector space $V$, a bilinear function $(x, y) \mapsto < x, y >$ (with values in the field of scalars) is called a symmetric inner product on V, whenever $< x, y > = < y, x >$ and is called an alternating inner product, whenever $< x, y > = -< y, x >$. A symmetric or an alternating inner product is called non-degenerate if, for every nonzero vector $x$, there exists a vector $y$ such that $< x, y > \neq 0$.

**Super spaces**: In a superspace $V = V_0 \oplus V_1$, the subspace $V_0$ is called the even part, and the subspace $V_1$ is called the odd part of $V$ [1][9]. Elements of $V_0$ are called even vectors, and elements of $V_1$ are called odd vectors and they are called homogeneous vectors of the space. We denote the parity of each homogeneous vector $x$ by $\varepsilon(x)$.

$$\varepsilon(x) = \begin{cases} 0 & x \text{ is even} \\ 1 & x \text{ is odd} \end{cases}$$

For simplicity in calculations, the value of $(-1)^{\varepsilon(x)}$ is shown by $(-1)^{x}$, and in this notation, the meaning of $x$ is the same as $\varepsilon(x)$. Whenever the symbol $\varepsilon(x)$ is used, it is assumed that $x$ is homogeneous. If the superspace $V$ has an algebra structure, we call it a super algebra whenever $\varepsilon(xy) = \varepsilon(x) + \varepsilon(y)$. (Addition is done in $\mathbb{Z}_2$)

A bilinear function on the super space $V$ is called super symmetric if the even and odd parts are perpendicular to each other and for homogeneous vectors we have:

$$< x, y >= (-1)^{xy} < y, x >$$

This condition shows that this operation on $V_0$ is a symmetric inner product and on $V_1$ is an alternating inner product. The non-degeneracy of this multiplication is equivalent to the non-degeneracy on $V_0$ and $V_1$.

**Algebra of alternating tensors:** Let $V$ be a vector space. the space of alternating tensors (contravariant) of order $k$ on $V$ is denoted by $\Lambda^{\mathrm{k}} V$, whose simple elements are $x_1 \wedge \cdots \wedge x_k$. The direct sum of these spaces forms a graded associative super commutative algebra by external product [4].

$$\Lambda V = \oplus_{k=0}^{n} \Lambda^k V \quad , n = \dim(V)$$

Considering a non-degenerate symmetric inner product on $V$, naturally, a non-degenerate symmetric inner product on $\Lambda^k V$ ($k = 0, \cdots, n$) is created as follows.

$$\begin{aligned} < x_1 \wedge \cdots \wedge x_k , y_1 \wedge \cdots \wedge y_k >&= \textstyle\sum_\sigma \varepsilon_\sigma < x_1, y_{\sigma(1)} > \cdots < x_k, y_{\sigma(k)} > \\ &= \det(< x_i, y_j >) \end{aligned} \tag{1}$$

This definition is well-defined and can be extended to the whole $\Lambda V$ by considering the spaces $\Lambda^k V$ to be perpendicular to each other. Elements of $\Lambda V$ are called multivectors and we denote them by $\boldsymbol{x}, \boldsymbol{y}, \boldsymbol{z}, \cdots$.

Some important involutions on $\Lambda V$ are defined as follows [4].

$$\begin{aligned} (x_1 \wedge \cdots \wedge x_k)' &= (-1)^k x_1 \wedge \cdots \wedge x_k \\ (x_1 \wedge \cdots \wedge x_k)^{\mathrm{opp}} &= x_{\mathrm{k}} \wedge \cdots \wedge x_1 = (-1)^{\frac{k(k-1)}{2}} x_1 \wedge \cdots \wedge x_k \\ \overline{x_1 \wedge \cdots \wedge x_k} &= (x_1 \wedge \cdots \wedge x_k)'^{\mathrm{opp}} = (-1)^{\frac{k(k+1)}{2}} x_1 \wedge \cdots \wedge x_k \end{aligned} \tag{2}$$

The following relations hold for these involutions.

$$(\boldsymbol{x} \wedge \boldsymbol{y})' = \boldsymbol{x}' \wedge \boldsymbol{y}' \ , (\boldsymbol{x} \wedge \boldsymbol{y})^{opp} = \boldsymbol{y}^{opp} \wedge \boldsymbol{x}^{opp} \ , \overline{\boldsymbol{x} \wedge \boldsymbol{y}} = \overline{\boldsymbol{y}} \wedge \overline{\boldsymbol{x}} \tag{3}$$

The insertion of a vector $v$ in the members of $\bigwedge V$ is an operator that is defined as follows.

$$i_v(x_1 \wedge \cdots \wedge x_k) = \sum_{j=1}^{k}(-1)^{j+1} < v, x_j > x_1 \wedge \cdots \wedge \widehat{x_j} \wedge \cdots \wedge x_k \tag{4}$$

This operator is a derivation of degree -1 on $\bigwedge V$. This operator has the following relations to the above involutions.

$$(i_v\boldsymbol{x})' = -i_v\boldsymbol{x}' \ , \ (i_v\boldsymbol{x})^{\text{opp}} = -i_v\overline{\boldsymbol{x}} \ , \qquad \overline{i_v\boldsymbol{x}} = i_v\boldsymbol{x}^{opp} \tag{5}$$

An important relation between the insertion operator and the symmetric inner product on $\bigwedge V$ is as follows [4].

$$< i_v\boldsymbol{x}, \boldsymbol{y} >=< \boldsymbol{x}, v \wedge \boldsymbol{y} > \tag{6}$$

The insertion operator is generalized for multivectors $\boldsymbol{x} = x_1 \wedge \cdots \wedge x_k$ and $\boldsymbol{y}$ by the following definition.

$$i_{\boldsymbol{x}}\boldsymbol{y} = i_{x_k} \circ \cdots \circ i_{x_1}(\boldsymbol{y}) \tag{7}$$

This definition is well-defined and if the order of $\boldsymbol{x}$ is strictly greater than the order of $\boldsymbol{y}$, then the value of the insertion will be zero. In general, the relationship between this operation and the symmetric inner product on $\bigwedge V$ is as follows.

$$< i_{\boldsymbol{x}}\boldsymbol{y}, \boldsymbol{z} >=< \boldsymbol{y}, \boldsymbol{x} \wedge \boldsymbol{z} > \tag{8}$$

The above relation shows that if $\boldsymbol{x}$ and $\boldsymbol{y}$ are of the same order, then:

$$i_{\boldsymbol{x}}\boldsymbol{y} =< \boldsymbol{x}, \boldsymbol{y} > \tag{9}$$

**Algebra of symmetric tensors**: The space of symmetric (contravariant) tensors of order $k$ on a vector space W is denoted by $\bigvee^{k} W$, and its simple elements are $\xi_1 \vee \cdots \vee \xi_k$ [4]. The direct sum of these spaces is denoted by $\bigvee W$ which is a commutative graded associative algebra by its symmetric product.

$$\bigvee W = \oplus_{k=0}^{\infty} \bigvee^k W$$

An important involution on $\bigvee W$ is the parity operator which is defined as follows.

$$(\xi_1 \vee \cdots \vee \xi_k)' = (-1)^k \xi_1 \vee \cdots \vee \xi_k$$

We consider a non-degenerate alternating inner product $\omega$ on $W$. So $W$ must be even dimensional. By $\omega$, we can construct a non-degenerate inner product on$\vee^k W$ as follows

$$< \xi_1 \vee \cdots \vee \xi_k , \eta_1 \vee \cdots \vee \eta_k >= \sum_{\sigma \in S_k} \omega(\xi_1, \eta_{\sigma(1)}) \cdots \omega(\xi_k, \eta_{\sigma(k)}) \quad (10)$$

By considering the subspaces $\vee^k W$ to be orthogonal to each other, we can extend these inner products on the whole $\vee W$. This inner product for even $k$, is symmetric, and for odd $k$, is alternating. Therefore, for the superspace structure of $\vee W$, this inner product is a supersymmetric inner product on this superalgebra.

Arbitrary elements of $\vee W$ will be denoted by $\boldsymbol{\xi}, \boldsymbol{\eta}, \boldsymbol{\phi}, \ldots,$ and are called symmetric multivectors. The insertion of a vector $\eta$ in the members of $\vee W$ is defined as follows.

$$i_\eta(\xi_1 \vee \cdots \vee \xi_k) = \sum_{j=1}^{k} \omega(\eta, \xi_j) \xi_1 \vee \cdots \vee \widehat{\xi_j} \vee \cdots \vee \xi_k \quad (11)$$

This operation is also a derivation of degree -1 on the algebra $\vee W$ The important relation between this operator and the supersymmetric inner product on $\vee W$ is as follows.

$$< i_\eta \boldsymbol{\xi}, \boldsymbol{\phi} >= -< \boldsymbol{\xi}, \eta \vee \boldsymbol{\phi} > \quad (12)$$

Note that in the above, one of the inner products, is symmetric and the other is alternating. This insertion operator is extended for $\boldsymbol{\eta} = \eta_1 \vee \cdots \vee \eta_k$ and $\boldsymbol{\xi}$ as follws.

$$i_{\boldsymbol{\eta}} \boldsymbol{\xi} = i_{\eta_1} \circ \cdots \circ i_{\eta_k}(\boldsymbol{\xi}) \quad (13)$$

If the order of $\boldsymbol{\eta}$ is strictly greater than the order of $\boldsymbol{\xi}$, then the value of the insertion is zero. In general, the relationship between this operation and the supersymmetric inner product is as follows.

$$< i_{\boldsymbol{\eta}} \boldsymbol{\xi}, \boldsymbol{\phi} >= (-1)^{\boldsymbol{\eta}} < \boldsymbol{\xi}, \boldsymbol{\eta} \vee \boldsymbol{\phi} > \quad (14)$$

The above relation shows that if $\boldsymbol{\eta}$ and $\boldsymbol{\xi}$ are of the same order, then:

$$i_{\boldsymbol{\eta}} \boldsymbol{\xi} = (-1)^{\boldsymbol{\eta}} < \boldsymbol{\xi}, \boldsymbol{\eta} >=< \boldsymbol{\eta}, \boldsymbol{\xi} > \quad (15)$$

2. **Structure of Clifford Algebra of a symmetric inner product**

In this section, $V$ is a vector space and $<,>$ is a non-degenerate symmetric inner product on $V$. For an associative unital algebra $A$, a linear map $\varphi: V \longrightarrow A$ is called a Clifford map whenever:

$$\varphi(x)\varphi(y) + \varphi(y)\varphi(x) = 2 < x, y > \mathbf{1} \tag{16}$$

There exists a unique associative unital algebra (up to isomorphism)[4][8], denoted by $CL(V)$, such that $V$ is a subspace of $CL(V)$ and generates it, and

$$\forall x, y \in V \qquad xy + yx = 2 < x, y > \mathbf{1} \tag{17}$$

and for every unital associative algebra $A$, every Clifford map $\varphi: V \longrightarrow A$ is extended uniquely to an algebra homomorphism $\tilde{\varphi}: CL(V) \longrightarrow A$. The algebra $CL(V)$ is called the Clifford algebra of $(V, <,>)$.

So, for any two orthogonal vectors $x, y$ we have $xy = -yx$. The algebra $CL(V)$ as a vector space, is naturally isomorphic to $\wedge V$ and the following map is the natural isomorphism between these spaces.

$$x_1 \wedge \cdots \wedge x_k \longmapsto \frac{1}{k!} \sum_{\sigma \in S_k} \varepsilon_\sigma x_{\sigma(1)} \cdots x_{\sigma(k)} \tag{18}$$

This definition is well defined and if $x_1, \cdots, x_k$ are prepenicular to each other then $x_1 \wedge \cdots \wedge x_k$ is mapped to $x_1 \cdots x_k$. Therefore, for an orthogonal basis $\{e_1, \cdots, e_n\}$ the basis that is created on $\wedge V$ is mapped to the basis that is created on $CL(V)$ so, the above map is an isomorphism. With this isomorphism, we can consider the exterior multiplication on $\wedge V$ as an operation on $CL(V)$. So, for two vectors $x, y \in V$ we have:

$$x \wedge y = \frac{1}{2}(xy - yx) = xy - < x, y > \mathbf{1} \;\Rightarrow\; xy = x \wedge y + < x, y > \mathbf{1} \tag{19}$$

Three involutions on $\wedge V$ that have been described above, can also be considered as involutions on $CL(V)$. and the following relations hold.

$$(\boldsymbol{xy})' = \boldsymbol{x}'\boldsymbol{y}' \;, (\boldsymbol{xy})^{opp} = \boldsymbol{y}^{opp}\boldsymbol{x}^{opp} \;, \overline{\boldsymbol{xy}} = \overline{\boldsymbol{y}}\,\overline{\boldsymbol{x}} \tag{20}$$

Clifford product between vectors in $V$ and multivectors in $CL(V)$ satisfy the following relations.

$$x(y_1 \wedge \cdots \wedge y_k) = x \wedge y_1 \wedge \cdots \wedge y_k + i_x(y_1 \wedge \cdots \wedge y_k) \tag{21}$$

$$(y_1 \wedge \cdots \wedge y_k)x = y_1 \wedge \cdots \wedge y_k \wedge x - (-1)^k i_x(y_1 \wedge \cdots \wedge y_k) \tag{22}$$

Due to the linearity of both sides of equalities for individual vectors, it is enough to check the correctness of equality for the vectors of an orthogonal base.

The involution $\boldsymbol{x} \longmapsto \boldsymbol{x}'$, makes both $CL(V)$ and $\wedge V$ into a superalgebra. A multivector $\boldsymbol{x}$ is even whenever $\boldsymbol{x}' = \boldsymbol{x}$ and is odd whenever $\boldsymbol{x}' = -\boldsymbol{x}$. We can consider Lie algebra structure and Lie superalgebra structure [6] on $CL(V)$ by the following Lie brackets.

$$\begin{array}{ll} \text{(Lie bracket)} & [\boldsymbol{x}, \boldsymbol{y}] = \boldsymbol{x}\boldsymbol{y} - \boldsymbol{y}\boldsymbol{x} \\ \text{(super Lie bracket)} & [\boldsymbol{x}, \boldsymbol{y}]^s = \boldsymbol{x}\boldsymbol{y} - (-1)^{xy}\boldsymbol{y}\,\boldsymbol{x} \end{array} \tag{23}$$

Note that if one of $\mathrm{x}, \mathrm{y}$ is even. then Lie bracket and super Lie bracket of $\mathrm{x}, \mathrm{y}$ are the same.

The Super Lie bracket operation between a vector $x$ and a multivector $\boldsymbol{y}$ is the same as two times the insertion $x$ into $\boldsymbol{y}$. In fact:

$$[x, \boldsymbol{y}]^s = 2i_x\boldsymbol{y} \tag{24}$$

To prove this equality, it is sufficient to use equalities (21) and (22).

To find the general formula for the Clifford multiplication of multivector, it is necessary to generalize the insertion operator such that it also includes the exterior multiplication. For every natural number $1 \leq k$ and a natural number $1 \leq l \leq k$ and a multivector $\boldsymbol{y}$, we put

$$i^{(l)}_{x_1 \wedge \cdots \wedge x_k}\boldsymbol{y} = \sum_{\sigma \in S_{l,k-l}} \varepsilon_\sigma x_{\sigma(l+1)} \wedge \cdots \wedge x_{\sigma(k)} \wedge i_{x_{\sigma(l)} \wedge \cdots \wedge x_{\sigma(1)}}(\boldsymbol{y}) \tag{25}$$

This definition is well defined. In the above:

$$S_{l,k-l} = \{\sigma \in S_k \mid \sigma(1) < \cdots < \sigma(l)\,, \sigma(l+1) < \cdots < \sigma(k)\ \} \tag{26}$$

In the case $l = 0$ it is natural to define $i^{(0)}_{\boldsymbol{x}}\boldsymbol{y} = \boldsymbol{x} \wedge \boldsymbol{y}$. If the order of $\boldsymbol{x}$ or $\boldsymbol{y}$ is strictly less than $l$, then $i^{(l)}_{\boldsymbol{x}}\boldsymbol{y} = 0$. If $\boldsymbol{y}$ is of order $m$, this formula shows that the order of $i^{(l)}_{\boldsymbol{x}}\boldsymbol{y}$ is equal to $k + m - 2l$.

It is clear that in the case of $l = k$, we have $i_{\boldsymbol{x}}^{(k)} = i_{\boldsymbol{x}^{opp}}$ Therefore, for two elements of the same order $k$ such as $\boldsymbol{x}$, $\boldsymbol{y}$ we also have:

$$i_{\boldsymbol{x}}^{(k)}\boldsymbol{y} = i_{\boldsymbol{x}^{opp}}\boldsymbol{y} =< \boldsymbol{x}^{opp}, \boldsymbol{y} >= (-1)^{\frac{k(k-1)}{2}} < \boldsymbol{x}, \boldsymbol{y} > \tag{27}$$

**Theorem 1**. The following relations hold for all multivectors $\boldsymbol{x}, \boldsymbol{y}$, and a vector $v$.

$$i_{\boldsymbol{x}}^{(l)}\boldsymbol{y} = (-1)^{(x-l)(y-l)} i_{\boldsymbol{y}}^{(l)}\boldsymbol{x} \tag{28}$$

$$i_{v\wedge\boldsymbol{x}}^{(l)}(\boldsymbol{y}) = (-1)^{l} v \wedge i_{\boldsymbol{x}}^{(l)}(\boldsymbol{y}) + (-1)^{l-1} i_{\boldsymbol{x}}^{(l-1)} \circ i_v(\boldsymbol{y}) \tag{29}$$

$$i_v \circ i_{\boldsymbol{x}}^{(l)}(\boldsymbol{y}) = (-1)^{l} i_{i_v(\boldsymbol{x})}^{(l)}(\boldsymbol{y}) + (-1)^{x} i_{\boldsymbol{x}}^{(l)} \circ i_v(\boldsymbol{y}) \tag{30}$$

$$i_{\boldsymbol{x}}^{(l)}(v \wedge \boldsymbol{y}) = (-1)^{x} v \wedge i_{\boldsymbol{x}}^{(l)}(\boldsymbol{y}) + (-1)^{l-1} i_{i_v(\boldsymbol{x})}^{(l-1)}(\boldsymbol{y}) \tag{31}$$

**Proof**: Due to the linearity of both sides of these equalities for their variables, it is enough to prove them for the members of a base. In particular, for an orthogonal base $\{e_1, \cdots, e_n\}$ of $V$, we can use the exterior products of vectors $e_1, \cdots, e_n$ to construct a base for $\wedge V$. Let $\boldsymbol{e}, \boldsymbol{u}, \boldsymbol{v}, \ldots$ be some multivectors in the base which is constructed for $\wedge V$. It is easy to see that if the number of identical vectors in $\boldsymbol{u}$ and $\boldsymbol{v}$ is not equal to $l$ then $i_{\boldsymbol{u}}^{(l)}(\boldsymbol{v}) = 0$. And if oreder of $\boldsymbol{e}$ is $l$, then:

$$i_{\boldsymbol{e}\wedge\boldsymbol{u}}^{(l)}\boldsymbol{e} \wedge \boldsymbol{v} = (-1)^{\frac{l(l-1)}{2}} \boldsymbol{u} \wedge \boldsymbol{v} \tag{32}$$

All the above equations in the theorem can be proved with the help of this relation by checking all possible situations ■

We also define two generalized insertion operators as follows. For a multivector $\boldsymbol{x}$ of order $k$ define:

$$\imath_{\boldsymbol{x}} = \sum_{l=0}^{k} i_{\boldsymbol{x}}^{(l)} \quad , \quad \imath'_{\boldsymbol{x}} = \sum_{l=0}^{k} (-1)^{l} i_{\boldsymbol{x}}^{(l)} \tag{33}$$

For example, for a vector $v$ and a multivector $\boldsymbol{y}$, we have:

$$\imath_v \boldsymbol{y} = v \wedge \boldsymbol{y} + i_v \boldsymbol{y} = v\boldsymbol{y} \quad . \quad \imath'_v \boldsymbol{y} = v \wedge \boldsymbol{y} - i_v \boldsymbol{y} = \boldsymbol{y}' v \tag{34}$$

By adding the equalities (29), (30), (31) for all possible values of $l$, the following similar equalities for the generalized insertion operators are obtained.

$$
\begin{aligned}
\imath_{v\wedge \boldsymbol{x}}(\boldsymbol{y}) &= v \wedge \imath'_{\boldsymbol{x}}(\boldsymbol{y}) + \imath'_{\boldsymbol{x}} \circ i_v(\boldsymbol{y}) \\
\imath'_{v\wedge \boldsymbol{x}}(\boldsymbol{y}) &= v \wedge \imath_{\boldsymbol{x}}(\boldsymbol{y}) - \imath_{\boldsymbol{x}} \circ i_v(\boldsymbol{y})
\end{aligned} \tag{35}
$$

$$
\begin{aligned}
i_v \circ \imath_{\boldsymbol{x}}(\boldsymbol{y}) &= \imath'_{i_v(x)}(\boldsymbol{y}) + (-1)^{\boldsymbol{x}} \imath_{\boldsymbol{x}} \circ i_v(\boldsymbol{y}) \\
i_v \circ \imath'_{\boldsymbol{x}}(\boldsymbol{y}) &= \imath_{i_v(x)}(\boldsymbol{y}) + (-1)^{\boldsymbol{x}} \imath'_{\boldsymbol{x}} \circ i_v(\boldsymbol{y})
\end{aligned} \tag{36}
$$

$$
\begin{aligned}
\imath_{\boldsymbol{x}}(v \wedge \boldsymbol{y}) &= (-1)^{\boldsymbol{x}} v \wedge \imath_{\boldsymbol{x}}(\boldsymbol{y}) + \imath'_{i_v(x)}(\boldsymbol{y}) \\
\imath'_{\boldsymbol{x}}(v \wedge \boldsymbol{y}) &= (-1)^{\boldsymbol{x}} v \wedge \imath'_{\boldsymbol{x}}(\boldsymbol{y}) - \imath_{i_v(x)}(\boldsymbol{y})
\end{aligned} \tag{37}
$$

Also, by adding the equalities (28) for all possible values of $l$, we find:

$$
\begin{cases}
\imath_{\boldsymbol{x}}\, \boldsymbol{y} = \imath'_{\boldsymbol{y}} \boldsymbol{x} & \text{both } \boldsymbol{x}, \boldsymbol{y} \text{ are even} \\
\imath_{\boldsymbol{x}}\, \boldsymbol{y} = -\imath'_{\boldsymbol{y}} \boldsymbol{x} & \text{both } \boldsymbol{x}, \boldsymbol{y} \text{ are odd}
\end{cases} \tag{38}
$$

Also, If one of the $\boldsymbol{x}, \boldsymbol{y}$ is even and the other is odd, then:

$$
\imath_{\boldsymbol{x}}\, \boldsymbol{y} = \imath_{\boldsymbol{y}}\, \boldsymbol{x}\ ,\ \ \imath'_{\boldsymbol{x}} \boldsymbol{y} = \imath'_{\boldsymbol{y}} \boldsymbol{x} \tag{39}
$$

**Theorem 2**: The Clifford multiplication and the generalized insertion operators are related as follows.

$$
\begin{cases}
\boldsymbol{x}\boldsymbol{y} = \imath_{\boldsymbol{x}}(\boldsymbol{y}) & \boldsymbol{x} \text{ is odd} \\
\boldsymbol{x}\boldsymbol{y} = \imath'_{\boldsymbol{x}}(\boldsymbol{y}) & \boldsymbol{x} \text{ is even}
\end{cases} \tag{40}
$$

**Proof**: We prove this theorem by induction on $k$ (order of $\boldsymbol{x}$). For $k=1$, we have already obtained this result in (34). For arbitrary $k$, we separate two cases of evenness and oddness of $k$. First, suppose (40) holds for all $\boldsymbol{x}$ of order $2m-1$ and we prove (40) for all $\boldsymbol{x}$ of order $2m$. It is sufficient to prove (40) for multivectors in the form $v \wedge x$, where $v$ is a vector and $\boldsymbol{x}$ is a multivector of order $2m-1$.

$$
\begin{aligned}
(v \wedge \boldsymbol{x})\boldsymbol{y} &= \tfrac{1}{2}(v\boldsymbol{x} - \boldsymbol{x}v)\boldsymbol{y} = \tfrac{1}{2}(v\boldsymbol{x}\boldsymbol{y} - \boldsymbol{x}v\boldsymbol{y}) = \tfrac{1}{2}\big(v(\imath_{\boldsymbol{x}}\boldsymbol{y}) - \boldsymbol{x}(v \wedge \boldsymbol{y} + i_v\boldsymbol{y})\big) \\
&= \tfrac{1}{2}\Big(v \wedge \imath_{\boldsymbol{x}}\boldsymbol{y} + i_v(\imath_{\boldsymbol{x}}\boldsymbol{y}) - \big(\imath_{\boldsymbol{x}}(v \wedge \boldsymbol{y}) + \imath_{\boldsymbol{x}}(i_v\boldsymbol{y})\big)\Big) \\
&= \tfrac{1}{2}\Big(v \wedge \imath_{\boldsymbol{x}}\boldsymbol{y} + \imath'_{i_v(x)}(\boldsymbol{y}) - \imath_{\boldsymbol{x}} \circ i_v(\boldsymbol{y}) - \big(-v \wedge \imath_{\boldsymbol{x}}(\boldsymbol{y}) + \imath'_{i_v(x)}(\boldsymbol{y}) + \imath_{\boldsymbol{x}}(i_v\boldsymbol{y})\big)\Big) \\
&= v \wedge \imath_{\boldsymbol{x}}\boldsymbol{y} - \imath_{\boldsymbol{x}} \circ i_v(\boldsymbol{y}) = \imath'_{v\wedge \boldsymbol{x}}(\boldsymbol{y})
\end{aligned} \tag{41}
$$

Now, suppose (40) holds for all $\boldsymbol{x}$ of order $2m$ and we prove (40) for all $\boldsymbol{x}$ of order $2m+1$. As before, we prove (40) for multivectors in the form $v \wedge \boldsymbol{x}$.

$$
\begin{aligned}
(v \wedge \boldsymbol{x})\boldsymbol{y} = \tfrac{1}{2}(v\boldsymbol{x} + \boldsymbol{x}v)\boldsymbol{y} = \tfrac{1}{2}(v\boldsymbol{x}\boldsymbol{y} + \boldsymbol{x}v\boldsymbol{y}) = \tfrac{1}{2}\big(v(\imath'_{\boldsymbol{x}}\boldsymbol{y}) + \boldsymbol{x}(v \wedge \boldsymbol{y} + i_v\boldsymbol{y})\big) \\
= \tfrac{1}{2}\big(v \wedge \imath'_{\boldsymbol{x}}\boldsymbol{y} + i_v(\imath'_{\boldsymbol{x}}\boldsymbol{y}) + \imath'_{\boldsymbol{x}}(v \wedge \boldsymbol{y}) + \imath'_{\boldsymbol{x}}(i_v\boldsymbol{y})\big) \\
= \tfrac{1}{2}\Big(v \wedge \imath'_{\boldsymbol{x}}\boldsymbol{y} + \imath_{i_v(\boldsymbol{x})}(\boldsymbol{y}) + \imath'_{\boldsymbol{x}} \circ i_v(\boldsymbol{y}) + v \wedge \imath'_{\boldsymbol{x}}(\boldsymbol{y}) - \imath_{i_v(\boldsymbol{x})}(\boldsymbol{y}) + \imath'_{\boldsymbol{x}}(i_v\boldsymbol{y})\Big) \\
= v \wedge \imath'_{\boldsymbol{x}}\boldsymbol{y} + \imath'_{\boldsymbol{x}} \circ i_v(\boldsymbol{y}) = \imath_{v\wedge\boldsymbol{x}}(\boldsymbol{y})
\end{aligned} \tag{42}
$$

So, the proof is complete. ■

**Cororally** 3:

$$
[\boldsymbol{x}, \boldsymbol{y}]^s = (-1)^{\boldsymbol{x}}\left(\imath'_{\boldsymbol{x}}(\boldsymbol{y}) - \imath_{\boldsymbol{x}}\ (\boldsymbol{y})\right) = (-1)^{\boldsymbol{x}+1} 2 \sum_{k=0}^{\infty} i_{\boldsymbol{x}}^{(2k+1)}\boldsymbol{y} \tag{43}
$$

To prove this equation, it can be checked for all possible cases of evenness and oddness of $\boldsymbol{x}, \boldsymbol{y}$. In the case $\boldsymbol{x} = u \wedge v$, we have:

$$
[u \wedge v, \boldsymbol{y}] = [u \wedge v, \boldsymbol{y}]^s = -2 i_{u\wedge v}^{(1)}\boldsymbol{y} = 2(u \wedge i_v\boldsymbol{y} - v \wedge i_u\boldsymbol{y}) \tag{44}
$$

In particular, when $\boldsymbol{y} = y_1 \wedge \cdots \wedge y_k$ we have :

$$
[u \wedge v, y_1 \wedge \cdots \wedge y_k] = \sum_{i=1}^{k} y_1 \wedge \cdots \wedge [u \wedge v, y_i] \wedge \cdots \wedge y_k \tag{45}
$$

Now, we will discuss the relations between the Clifford product and Lie bracket and the symmetric inner product on $CL(V)$.

**Theorem** 4**:** For a multivector $\boldsymbol{x}$ of order $k$ and any multivectors $\boldsymbol{y}, \boldsymbol{z}$ and a number $0 \le l \le k$ we have:

$$
< i_{\boldsymbol{x}}^{(l)}\boldsymbol{y}, \boldsymbol{z} >=< \boldsymbol{y}, i_{\boldsymbol{x}^{opp}}^{(k-l)}\boldsymbol{z} > \tag{46}
$$

**Proof**: The only important case is when the order of the multivectors fits together, in other cases both sides are zero. Therefore, If the order of $\boldsymbol{y}$ is $m$ , then the order of $\boldsymbol{z}$ must be equal to $k + m - 2l$. We prove the theorem by induction on $k$.

For $k = 1$ and $l = 1$, (46) is the same as (6). By interchanging the sides of (6) we find (46) for $l = 0$.

Now, suppose that equality holds for all multivectors of order $k$. We prove the equality for all multivectors in the form $u \wedge \boldsymbol{x}$ where $\boldsymbol{x}$ is of order k.

$$
\begin{aligned}
< i_{u\wedge x}^{(l)}\boldsymbol{y},\boldsymbol{z} > &= < (-1)^l u \wedge i_{\boldsymbol{x}}^{(l)}\boldsymbol{y} + (-1)^{l-1} i_{\boldsymbol{x}}^{(l-1)} \circ i_u(\boldsymbol{y}), \boldsymbol{z} > \\
&= (-1)^l < i_{\boldsymbol{x}}^{(l)}\boldsymbol{y}, i_u\boldsymbol{z} > +(-1)^{l-1} < i_u(\boldsymbol{y}), i_{\boldsymbol{x}^{opp}}^{(k-l+1)}\boldsymbol{z} > \\
&= < \boldsymbol{y}, (-1)^l i_{\boldsymbol{x}^{opp}}^{(k-l)} \circ i_u(\boldsymbol{z}) + (-1)^{l-1} u \wedge i_{\boldsymbol{x}^{opp}}^{(k-l+1)}\boldsymbol{z} > \\
&= (-1)^k < \boldsymbol{y}, i_{u\wedge \boldsymbol{x}^{opp}}^{(k+1-l)}\boldsymbol{z} > = < \boldsymbol{y}, i_{(u\wedge \boldsymbol{x})^{opp}}^{(k+1-l)}\boldsymbol{z} > \quad \blacksquare
\end{aligned}
\tag{47}
$$

By summation of the sides of (46) for all possible values of $l$, we obtain the following results.

$$
\begin{gathered}
< i_{\boldsymbol{x}}\, \boldsymbol{y}, \boldsymbol{z} > = < \boldsymbol{y}, i_{\boldsymbol{x}^{opp}}\boldsymbol{z} > \\
< i'_{\boldsymbol{x}}\boldsymbol{y}, \boldsymbol{z} > = < \boldsymbol{y}, i'_{\bar{\boldsymbol{x}}}\boldsymbol{z} >
\end{gathered}
\tag{48}
$$

**Cororally** 5: For all multivectors $\boldsymbol{x}, \boldsymbol{y}, \boldsymbol{z}$ we have:

$$
< \boldsymbol{xy}, \boldsymbol{z} > = < \boldsymbol{x}, \boldsymbol{z}\boldsymbol{y}^{opp} > = < \boldsymbol{y}, \boldsymbol{x}^{opp}\boldsymbol{z} > \tag{49}
$$

This equality can be checked for all possible cases of evenness and oddness of $\boldsymbol{x}, \boldsymbol{y}, \boldsymbol{z}$. Note that the only important cases are $\varepsilon(\boldsymbol{x}) + \varepsilon(\boldsymbol{y}) + \varepsilon(\boldsymbol{z}) = 0$. In other cases, both sides of (48) are zero.

**Cororally** 6. For Lie algebra structure of $CL(V)$, we have:

$$
< [\boldsymbol{x}, \boldsymbol{y}], \boldsymbol{z} > = < \boldsymbol{x}, [\boldsymbol{z}, \boldsymbol{y}^{opp}] > \tag{50}
$$

To make an inner product that is invariant under the Lie algebra structure of $CL(V)$, we need a slight change in the inner product. We define a new inner product on $\Lambda V$ as follows.

$$
< \boldsymbol{x}, \boldsymbol{y} >_{new} = < \boldsymbol{x}^{opp}, \boldsymbol{y} > \tag{51}
$$

This new symmetric inner product may be different from the previous one in at most a sign, and with this new inner product, the following relations hold.

$$
< \boldsymbol{xy}, \boldsymbol{z} >_{new} = < \boldsymbol{x}, \boldsymbol{yz} >_{new} = < \boldsymbol{y}, \boldsymbol{zx} >_{new} \tag{52}
$$

$$
< [\boldsymbol{x}, \boldsymbol{y}\,], \boldsymbol{z} >_{new} = < \boldsymbol{x}, [\boldsymbol{y}, \boldsymbol{z}\,] >_{new} \tag{53}
$$

For each multivector $\boldsymbol{y}$, the operator $ad(\boldsymbol{y})$ on the $CL(V)$ is defined by $ad(\boldsymbol{y})(\boldsymbol{x}) = [\boldsymbol{y}, \boldsymbol{x}]$. If the order of $\boldsymbol{y}$ is $2$, then all subspaces $\Lambda^{\mathrm{k}}V$ are invariant under the operator $ad(\boldsymbol{y})$ and $ad(\boldsymbol{y})$ is antisymmetric for $<,>$ and $<,>_{new}$ and we have:

$$ad(\boldsymbol{y})(\boldsymbol{x}) = -2i_y^{(1)}\boldsymbol{x} \tag{54}$$

So, $\Lambda^2 V$ is closed under the Lie bracket and it is a Lie subalgebra $CL(V)$. In fact, all the subspaces $\Lambda^k V$ can be considered as a representation space for Lie algebra $\Lambda^2 V$.

The mapping $\boldsymbol{y} \longmapsto \mathrm{ad}(\boldsymbol{y})|_V$ is a Lie algebra isomorphism from $\Lambda^2 V$ to $\mathfrak{o}(V)$. Therefore, we can consider $\mathfrak{o}(V)$ as a Lie subalgebra of $CL(V)$. In particular, for each $T \in \mathfrak{o}(V)$ there is a unique element $\boldsymbol{y} \in \Lambda^2 V$ such that for each $\in V$ :

$$T(v) = [\boldsymbol{y}, v] = -2i_v\boldsymbol{y} \tag{55}$$

## 3. Structure of Weyl Algebra of an alternating inner product

In this section, $V$ is a vector space and $\omega$ is a non-degenerate alternating inner product on $V$ (So, $V$ is even dimensional). For an associative unital algebra $A$, a linear map $\varphi: V \longrightarrow A$ is called a Weyl map whenever:

$$\varphi(\xi)\varphi(\eta) - \varphi(\eta)\varphi(\xi) = 2\omega(\xi,\eta)\mathbf{1} \tag{56}$$

There exists a unique associative unital algebra (up to isomorphism)[10], denoted by $WL(V)$, such that $V$ is a subspace of $WL(V)$ and generates it, and

$$\forall \xi, \eta \in V \qquad \xi\eta - \eta\xi = 2\omega(\xi,\eta)\mathbf{1} \tag{57}$$

and for every unital associative algebra $A$, every Weyl map $\varphi: V \longrightarrow A$ is extended uniquely to an algebra homomorphism $\tilde{\varphi}: WL(V) \longrightarrow A$. The algebra $WL(V)$ is called the Weyl algebra of $(V, \omega)$.

So, for any two orthogonal vectors $\xi, \eta$ we have $\xi\eta = \eta\xi$. The map $\xi \mapsto -\xi$ from $V$ to $WL(V)$ is a Weyl map and extends to an algebra isomorphism on $WL(V)$. This is an involution on $WL(V)$ and its eigenspaces make the Weyl algebra into a superalgebra.

The algebra $WL(V)$ as a vector space, is naturally isomorphic to $\vee V$ and the following map is the natural isomorphism between these spaces.

$$\xi_1 \vee \cdots \vee \xi_k \longmapsto \frac{1}{k!}\sum_{\sigma\in S_k} \xi_{\sigma(1)} \cdots \xi_{\sigma(k)} \tag{58}$$

This definition is well-defined and if $\xi_1, \cdots, \xi_k$ are in an isotropic subspace then $\xi_1 \vee \cdots \vee \xi_k$ is mapped to $\xi_1 \cdots \xi_k$ . This isomorphism preserves the parity involutions of these two algebras and is an isomorphism of superalgebras. With this isomorphism, we can consider $\vee V$ and $WL(V)$ identical and the symmetric multiplication on $\vee V$ as an operation on $WL(V)$, or vice versa.

For example, for two vectors $\xi, \eta \in V$ we have:

$$\xi \vee \eta = \frac{1}{2}(\xi\eta + \eta\xi) = \xi\eta - \omega(\xi,\eta)\mathbf{1} \quad \Rightarrow \quad \xi\eta = \xi \vee \eta + \omega(\xi,\eta)\mathbf{1} \tag{59}$$

We denote the Lie bracket and the super Lie structure on $WL(V)$, respectively by $[,]$ and $[,]^s$.

**Theorem** 7. In the Weyl algebras, the Lie bracket operation between a vector and a symmetric multi-vector is the same as two times the insertion operation. In fact:

$$[\xi, \eta_1 \vee \cdots \vee \eta_k] = 2i_\zeta(\eta_1 \vee \cdots \vee \eta_k) \tag{60}$$

**Proof:**

$$\begin{aligned}
[\xi,\eta_1 \vee \cdots \vee \eta_k] &= \frac{1}{k!}\Sigma_\sigma \left[\xi,\eta_{\sigma(1)} \cdots \eta_{\sigma(k)}\right] \\
&= \frac{1}{k!}\Sigma_\sigma \left[\xi,\eta_{\sigma(1)}\right]\eta_{\sigma(2)} \cdots \eta_{\sigma(k)} + \cdots + \frac{1}{k!}\Sigma_\sigma\, \eta_{\sigma(1)} \cdots \eta_{\sigma(k-1)}\left[\xi,\eta_{\sigma(k)}\right] \\
&= \frac{1}{k!}\Sigma_\sigma\, 2\omega(\xi,\eta_{\sigma(1)})\eta_{\sigma(2)} \cdots \eta_{\sigma(k)} + \cdots + \frac{1}{k!}\Sigma_\sigma\, 2\omega(\xi,\eta_{\sigma(k)})\eta_{\sigma(1)} \cdots \eta_{\sigma(k-1)}
\end{aligned} \tag{61}$$

For any $j = 1, \cdots, k$ we can prove the following equality:

$$\begin{aligned}
\frac{1}{k!}\Sigma_\sigma\, 2\omega\big(\xi,\eta_{\sigma(j)}\big)\eta_{\sigma(1)} \cdots \hat{\eta}_{\sigma(j)} \cdots \eta_{\sigma(k)} &= \frac{2}{k}\Sigma_{i=1}^{k}\, \omega(\xi,\eta_i)\eta_1 \vee \cdots \vee \widehat{\eta_\iota} \vee \cdots \vee \eta_k \\
&= \frac{2}{k} i_\xi(\eta_1 \vee \cdots \vee \eta_k)
\end{aligned} \tag{62}$$

It is sufficient to note that for any $i = 1, \cdots, k$ we have:

$$\frac{1}{k!}\Sigma_{\sigma:\, \sigma(j)=i}\, 2\omega\big(\xi,\eta_{\sigma(j)}\big)\eta_{\sigma(1)} \cdots \hat{\eta}_{\sigma(j)} \cdots \eta_{\sigma(k)} = \frac{2}{k}\omega(\xi,\eta_i)\eta_1 \vee \cdots \vee \widehat{\eta_\iota} \vee \cdots \vee \eta_k \tag{63}$$

Now, the proof is complete. ■

**Theorem** 8. In the Weyl algebras, the multiplication of a vector by a symmetric multivector has the following relation to the symmetric product and the insertion operator.

$$\xi(\eta_1 \vee \cdots \vee \eta_k) = \xi \vee \eta_1 \vee \cdots \vee \eta_k + i_\xi(\eta_1 \vee \cdots \vee \eta_k) \tag{64}$$

**Proof**:

$$\begin{aligned} \xi \vee \eta_1 \vee \cdots \vee \eta_k &= \frac{1}{(k+1)!}\sum_\sigma \xi\eta_{\sigma(1)} \cdots \eta_{\sigma(k)} \\ &+ \cdots + \frac{1}{(k+1)!}\sum_\sigma \eta_{\sigma(1)} \cdots \eta_{\sigma(i)}\xi\eta_{\sigma(i+1)} \cdots \eta_{\sigma(k)} + \cdots + \frac{1}{(k+1)!}\sum_\sigma \eta_{\sigma(1)} \cdots \eta_{\sigma(k)}\xi \end{aligned} \tag{65}$$

First, we find the following relation for any terms in the above expression.

$$\begin{aligned} \frac{1}{(k+1)!}\sum_\sigma \eta_{\sigma(1)} \cdots \eta_{\sigma(i)}\xi\eta_{\sigma(i+1)} \cdots \eta_{\sigma(k)} &= \frac{1}{(k+1)!}\sum_\sigma \eta_{\sigma(1)} \cdots \eta_{\sigma(i-1)}\xi\eta_{\sigma(i)} \cdots \eta_{\sigma(k)} \\ &+ \frac{2}{(k+1)!}\sum_\sigma \omega\left(\eta_{\sigma(i)}, \xi\right)\eta_{\sigma(1)} \cdots \widehat{\eta}_{\sigma(i)} \cdots \eta_{\sigma(k)} \end{aligned} \tag{66}$$

The last term, by (62) is equal to $\frac{-2}{k(k+1)} i_\xi(\eta_1 \vee \cdots \vee \eta_k)$. Now, in (65), starting with the last term and using (66), and iterating this process we find (64). ■

**Cororally** 9.

$$\xi \vee \boldsymbol{\eta} = \frac{1}{2}(\xi\boldsymbol{\eta} + \boldsymbol{\eta}\xi) \quad , \ \boldsymbol{\eta}\xi = \boldsymbol{\eta} \vee \xi - i_\xi\boldsymbol{\eta} \,. \tag{67}$$

To find the general formula for the Weyl production of symmetric multivectors, it is necessary to generalize the insertion operator. For every natural number $1 \leq k$ and every natural number $l = 1, \cdots, k$, we put:

$$i^{(l)}_{\xi_1 \vee \cdots \vee \xi_k}\boldsymbol{\eta} = \sum_{\sigma \in S_{l,k-l}} \left(i_{\xi_{\sigma(1)} \vee \cdots \vee \xi_{\sigma(l)}}\boldsymbol{\eta}\right) \vee \left(\xi_{\sigma(l+1)} \vee \cdots \vee \xi_{\sigma(k)}\right) \tag{68}$$

In the case $l = 0$ , put $i^{(0)}_{\boldsymbol{\xi}}\boldsymbol{\eta} = \boldsymbol{\xi} \vee \boldsymbol{\eta}$ .

Explicitly, this operation of insertion between symmetric multivectors can be written as follows.

$$i^{(l)}_{\xi_1\vee\cdots\vee\xi_k}(\eta_1\vee\cdots\vee\eta_m) = \sum_{\sigma\in S_{l,k}\,\tau\in S_{l,m}}\omega(\xi_{\sigma(1)},\eta_{\tau(1)})\cdots\omega(\xi_{\sigma(l)},\eta_{\tau(l)})\,\xi_{\sigma(l+1)}\vee\cdots\vee\xi_{\sigma(k)}\vee\eta_{\tau(l+1)}\vee\cdots\vee\eta_{\tau(m)} \tag{69}$$

If the order of one of $\boldsymbol{\xi}$ or $\boldsymbol{\eta}$ is strictly less than $l$, then $i^{(l)}_{\boldsymbol{\xi}}\boldsymbol{\eta}=0$ This formula shows that the order of $i^{(l)}_{\boldsymbol{\xi}}\boldsymbol{\eta}$ is equal to $k+m-2l$. It is clear that in the case $l=k$ we have $i^{(k)}_{\boldsymbol{\xi}}=i_{\boldsymbol{\xi}}$. We also have: $i^{(l)}_{\boldsymbol{\xi}}\boldsymbol{\eta}=(-1)^l i^{(l)}_{\boldsymbol{\eta}}\boldsymbol{\xi}$ . From (69), the following equalities can also be easily proved:

$$i^{(l)}_{\xi\vee\boldsymbol{\eta}}(\boldsymbol{\phi})=\xi\vee i^{(l)}_{\boldsymbol{\eta}}(\boldsymbol{\phi})+i^{(l-1)}_{\boldsymbol{\eta}}\circ i_\xi(\boldsymbol{\phi}) \tag{70}$$

$$i_\xi\circ i^{(l)}_{\boldsymbol{\eta}}(\boldsymbol{\phi})=i^{(l)}_{i_\xi(\boldsymbol{\eta})}(\boldsymbol{\phi})+i^{(l)}_{\boldsymbol{\eta}}\circ i_\xi(\boldsymbol{\phi}) \tag{71}$$

$$i^{(l)}_{\boldsymbol{\eta}}(\xi\vee\boldsymbol{\phi})=\xi\vee i^{(l)}_{\boldsymbol{\eta}}(\boldsymbol{\phi})-i^{(l-1)}_{i_\xi\boldsymbol{\eta}}(\boldsymbol{\phi}) \tag{72}$$

In the context of the Weyl algebras, we also define two generalized insertion operators as follows. For a symmetric multivector $\boldsymbol{\xi}$ of order $k$ define:

$$\mathcal{i}_\xi=\sum_{l=0}^{k} i^{(l)}_\xi \quad , \quad \mathcal{i}'_\xi=\sum_{l=0}^{k}(-1)^l i^{(l)}_\xi \tag{73}$$

Clearly: $\mathcal{i}_\xi(\boldsymbol{\eta})=\mathcal{i}'_{\boldsymbol{\eta}}(\boldsymbol{\xi})$ , and for a vector $\xi$ we have:

$$\mathcal{i}_\xi\,\boldsymbol{\eta}=\xi\vee\boldsymbol{\eta}+i_\xi\boldsymbol{\eta}=\xi\boldsymbol{\eta} \tag{74}$$

By adding the sides of (70) and (71) and (72) for all possible values of $l$ we find similar relations for the generalized insertion operator.

$$\mathcal{i}_{\xi\vee\boldsymbol{\eta}}(\boldsymbol{\phi})=\xi\vee\mathcal{i}_{\boldsymbol{\eta}}(\boldsymbol{\phi})+\mathcal{i}_{\boldsymbol{\eta}}\circ i_\xi(\boldsymbol{\phi}) \tag{75}$$

$$i_\xi\circ\mathcal{i}_{\boldsymbol{\eta}}(\boldsymbol{\phi})=\mathcal{i}_{i_\xi(\boldsymbol{\eta})}(\boldsymbol{\phi})+\mathcal{i}_{\boldsymbol{\eta}}\circ i_\xi(\boldsymbol{\phi}) \tag{76}$$

$$\mathcal{i}_{\boldsymbol{\eta}}(\xi\vee\boldsymbol{\phi})=\xi\vee\mathcal{i}_{\boldsymbol{\eta}}(\boldsymbol{\phi})-\mathcal{i}_{i_\xi\boldsymbol{\eta}}(\boldsymbol{\phi}) \tag{77}$$

**Theorem** 10. Weyl multiplication in $WL(V)$ in terms of the generalized insertion operator is as follows.

$$\boldsymbol{\eta}\boldsymbol{\phi}=\mathcal{i}_{\boldsymbol{\eta}}(\boldsymbol{\phi})=\mathcal{i}'_{\boldsymbol{\phi}}\boldsymbol{\eta} \tag{78}$$

**Proof**: By inductionon $k$ (the order of $\boldsymbol{\eta}$), we can prove (78). For $k = 1$, (78) has already been obtained in Theorem 8. Now, suppose (78) holds for all $\boldsymbol{\eta}$ of order $k$, and we prove it for all symmetric multivectors in the form $\xi \vee \boldsymbol{\eta}$.

$$\begin{aligned}(\xi \vee \boldsymbol{\eta})\boldsymbol{\phi} &= \tfrac{1}{2}(\xi\boldsymbol{\eta} + \boldsymbol{\eta}\xi)\boldsymbol{\phi} = \tfrac{1}{2}(\xi\boldsymbol{\eta}\boldsymbol{\phi} + \boldsymbol{\eta}\xi\boldsymbol{\phi}) = \tfrac{1}{2}\big(\xi(i_{\boldsymbol{\eta}}\boldsymbol{\phi}) + \boldsymbol{\eta}(\xi \vee \boldsymbol{\phi} + i_\xi\boldsymbol{\phi})\big) \\ &= \tfrac{1}{2}\big(\xi \vee i_{\boldsymbol{\eta}}\boldsymbol{\phi} + i_\xi(i_{\boldsymbol{\eta}}\boldsymbol{\phi}) + i_{\boldsymbol{\eta}}(\xi \vee \boldsymbol{\phi}) + i_{\boldsymbol{\eta}}(i_\xi\boldsymbol{\phi})\big) \\ &= \tfrac{1}{2}\big(\xi \vee i_{\boldsymbol{\eta}}\boldsymbol{\phi} + i_{i_\xi(\boldsymbol{\eta})}(\boldsymbol{\phi}) + i_{\boldsymbol{\eta}} \circ i_\xi(\boldsymbol{\phi}) + \xi \vee i_{\boldsymbol{\eta}}(\boldsymbol{\phi}) - i_{i_\xi(\boldsymbol{\eta})}(\boldsymbol{\phi}) + i_{\boldsymbol{\eta}}(i_\xi\boldsymbol{\phi})\big) \\ &= \xi \vee i_{\boldsymbol{\eta}}(\boldsymbol{\phi}) + i_{\boldsymbol{\eta}} \circ i_\xi(\boldsymbol{\phi}) = i_{\xi\vee\boldsymbol{\eta}}(\boldsymbol{\phi}) \qquad \blacksquare\end{aligned} \tag{79}$$

**Cororally** 11. The lie bracket of symmetric multivectors satisfies the following relation.

$$[\boldsymbol{\xi}, \boldsymbol{\eta}] = i_{\boldsymbol{\xi}}(\boldsymbol{\eta}) - i_{\boldsymbol{\eta}}(\boldsymbol{\xi}) = i_\xi(\boldsymbol{\eta}) - i'_\xi(\boldsymbol{\eta}) = 2\sum_{n=0}^{\infty} i_\xi^{(2n+1)}\boldsymbol{\eta} \tag{80}$$

For example if order of $\boldsymbol{\xi}$ is 2 , then

$$[\boldsymbol{\xi}, \boldsymbol{\eta}]^s = [\boldsymbol{\xi}, \boldsymbol{\eta}] = 2i_\xi^{(1)}(\boldsymbol{\eta}) \tag{81}$$

Now, we explore the relationships between the super inner product and the Weyl multiplication and the super Lie bracket on $WL(V)$.

**Theorem** 12.For a symmetric multivector $\boldsymbol{\xi}$ of order $k$ and a number $0 \leq l \leq k$ we have:

$$< i_\xi^{(l)}\boldsymbol{\eta}, \boldsymbol{\phi} >= (-1)^l < \boldsymbol{\eta}, i_\xi^{(k-l)}\boldsymbol{\phi} > \tag{82}$$

**Proof**: The only important case is when the order of the multivectors are proportional, in other cases, both sides are zero. For example, if the order of $\boldsymbol{\eta}$ is $m$, the order of $\boldsymbol{\phi}$ must be equal to $k + m - 2l$. We prove the theorem by induction on $k$.

For $k = 1$ and $l = 1$ this equality is the same as (12). By interchanging the two terms in the inner product in (12), the theorem is proved for the case $l = 0$.

Now, suppose equality holds for all multivectors of order $k$ and we prove the equality for multivectors in the form $\gamma \vee \boldsymbol{\xi}$ where $\boldsymbol{\xi}$ is of order $k$.

$$
\begin{aligned}
< i_{\gamma\vee\xi}^{(l)}\boldsymbol{\eta},\boldsymbol{\phi} >&=< \gamma\vee i_{\xi}^{(l)}\boldsymbol{\eta} + i_{\xi}^{(l-1)}\circ i_{\gamma}(\boldsymbol{\eta}),\boldsymbol{\phi} > \\
&=< i_{\xi}^{(l)}\boldsymbol{\eta}, i_{\gamma}\boldsymbol{\phi} > +(-1)^{l-1} < i_{\gamma}(\boldsymbol{\eta}), i_{\xi}^{(k-l+1)}\boldsymbol{\phi} > \\
&= (-1)^{l} < \boldsymbol{\eta}, i_{\xi}^{(k-l)}\circ i_{\gamma}(\boldsymbol{\phi}) + \gamma\vee i_{\xi}^{(k-l+1)}\boldsymbol{\phi} > \\
&= (-1)^{l} < \boldsymbol{\eta}\,, i_{\gamma\vee\xi}^{(k+1-l)}\boldsymbol{\phi} > \qquad \blacksquare
\end{aligned} \tag{83}
$$

**Cororally** 13. If the order of $\boldsymbol{\xi}$ is , then:

$$< \boldsymbol{\xi\eta},\boldsymbol{\phi} >= (-1)^{k} < \boldsymbol{\eta},\boldsymbol{\phi\xi} > \tag{84}$$

it is sufficient to sum two sides of (82) for all possible values of $l$ and use(78).

**Cororally** 14.

$$< \boldsymbol{\xi\eta},\boldsymbol{\phi} >=< \boldsymbol{\xi},\boldsymbol{\eta\phi} > \tag{85}$$

Notice that to prove (85), the only important case is $\varepsilon(\boldsymbol{\xi}) + \varepsilon(\boldsymbol{\eta}) + \varepsilon(\boldsymbol{\phi}) = 0$, in other cases both sides are zero. To prove (85), interchange terms in the inner products in (84) and notice that for even $k$, these inner products on both sides are alternating or symmetric, and for odd $k$, one of them is alternating and the other is symmetric.

**Cororally** 15. The supersymmetric inner product on $WL(V)$ is invariant for the Lie superalgebra structure of the $WL(V)$. In other words:

$$< [\boldsymbol{\xi},\boldsymbol{\eta}]^{s},\boldsymbol{\phi} >=< \boldsymbol{\xi},[\boldsymbol{\eta},\boldsymbol{\phi}]^{s} > \tag{86}$$

Here, also the only important case is $\varepsilon(\boldsymbol{\xi}) + \varepsilon(\boldsymbol{\eta}) + \varepsilon(\boldsymbol{\phi}) = 0$. So, using (85), we can easily prove (86).

For every $\boldsymbol{\xi}$ define $\mathrm{ad}(\boldsymbol{\xi})\colon WL(V) \longrightarrow WL(V)$ by $\mathrm{ad}(\boldsymbol{\xi})(\boldsymbol{\eta}) = [\boldsymbol{\xi},\boldsymbol{\eta}]^{s}$ . If the order of $\boldsymbol{\xi}$ is 2, then:

$$\mathrm{ad}(\boldsymbol{\xi})(\boldsymbol{\eta}) = [\boldsymbol{\xi},\boldsymbol{\eta}]^{s} = [\boldsymbol{\xi},\boldsymbol{\eta}] = 2i_{\xi}^{(1)}\boldsymbol{\eta} \tag{87}$$

This formula shows that all subspaces $\vee^{k}V$ are invariant under $\mathrm{ad}(\boldsymbol{\xi})$, especially $\vee^{2}V$ is a Lie subalgebra of $WL(V)$ and all $\vee^{k}V$ are a representation for Lie algebra $\vee^{2}V$, and their inner product is invariant under this representation. Especially, $\mathrm{ad}(\boldsymbol{\xi})|_V \in \mathfrak{sp}(V)$ and the mapping $\boldsymbol{\xi} \longmapsto \mathrm{ad}(\boldsymbol{\xi})|_V$ is a Lie algebra isomorphism from $\bigvee^2 V$ to $\mathfrak{sp}(V)$. Therefore, we can consider $\mathfrak{sp}(V)$ as a Lie subalgebra of $WL(V)$, and for any $T \in \mathfrak{sp}(V)$, there exists a unique $\boldsymbol{\xi} \in \bigvee^2 V$ such that:

$$T(\eta) = [\boldsymbol{\xi}, \eta] = 2i_{\xi}^{(1)}\eta = -2i_{\eta}\boldsymbol{\xi} \tag{88}$$

## 4. Structure of Clifford-Weyl Algebra of a supersymmetric inner product

Now, we can combine Clifford and Weyl algebra and make a superalgebra that is called Clifford-Weyl algebra [5]. We must consider a super space $V = V_0 \oplus V_1$ and a supersymmetric inner product on it. For an associative unital algebra $A$, a linear map $\varphi: V \longrightarrow A$ is called a Clifford-Weyl map whenever for all homogeneous vectors $x, y \in V$:

$$\varphi(x)\varphi(y) + (-1)^{xy}\varphi(y)\varphi(x) = 2 < x, y > \mathbf{1} \tag{89}$$

In a superspace $V = V_0 \oplus V_1$ with a supersymmetric inner product, there exists a unique associative unital algebra, denoted by $CLW(V)$ which has the following properties.

1) $V$ is a subspace of $CLW(V)$ and is a generator of this algebra.

2) For all homogeneous elements $x, y \in V$ we have:

$$xy + (-1)^{xy}yx = 2 < x, y > \mathbf{1} \tag{90}$$

3) For every associative unital algebra $A$, any Clifford-Weyl map $\varphi: V \longrightarrow A$ is extended uniquely to an algebra homomorphism $\tilde{\varphi}: CLW(V) \longrightarrow A$.

This algebra naturally contains $CL(V_0)$and $WL(V_1)$. In fact, the inclusion maps $V_0 \longrightarrow CLW(V)$ and $V_1 \longrightarrow CLW(V)$ are respectively Clifford and Weyl maps, and extend to the following algebra homomorphism.

$$CL(V_0) \longrightarrow CLW(V)\,,\ WL(V_1) \longrightarrow CLW(V) \tag{91}$$

These homomorphisms imbed the algebras $CL(V_0)$and $WL(V_1)$ into $CLW(V)$. It can be proved that $CLW(V) = \mathrm{CL}(V_0)\widehat{\otimes}\mathrm{WL}(V_1)$ (tensor multiplication of super algebras [4]). It is enough to consider the following mapping which is a Clifford-Weyl map.

$$\begin{aligned} \mathrm{V}_0 \oplus \mathrm{V}_1 &\longrightarrow CL(\mathrm{V}_0)\widehat{\otimes} W\mathrm{L}(\mathrm{V}_1) \\ x+\xi &\longmapsto x\otimes 1 + 1\otimes\xi \end{aligned} \tag{92}$$

This map extends to an algebra homomorphism from $CLW(V)$ to $\mathrm{C}L(V_0)\widehat{\otimes} WL(V_1)$ which is an isomorphism.

For simplicity, we denote arbitrary elements of $\mathrm{C}L(V_0)$ with symbols $\boldsymbol{x}, \boldsymbol{y}, \boldsymbol{z}, \cdots$and elements of $V_0$ with symbols $x, y, z, \cdots$. Also, we denote arbitrary elements of $WL(V_1)$with the symbols $\boldsymbol{\xi}, \boldsymbol{\eta}, \boldsymbol{\phi}, \cdots$and elements of $V_1$ with the symbols $\xi, \eta . \phi, \cdots$

Even and odd vectors of $V$ are anticommutative. In other words, for $x \in V_0$and $\xi \in V_1$ we have $x\xi = -\xi x$. In general, $\boldsymbol{x\xi} = (-1)^{x\xi}\boldsymbol{\xi x}$. Simple elements of $CLW(V)$ are in the form $\boldsymbol{x} \otimes \boldsymbol{\xi}$.

The Lie algebra structure of $CLW(V)$ is related to the Lie algebra structures of $CL(V_0)$ and $WL(V_1)$ as follows.

$$\begin{aligned} [\boldsymbol{x}\otimes\boldsymbol{\xi}\ ,\boldsymbol{y}\otimes\boldsymbol{\eta}\ ] &= [\boldsymbol{x},\boldsymbol{y}]\otimes\boldsymbol{\xi\eta} + \boldsymbol{yx}\otimes[\boldsymbol{\xi},\boldsymbol{\eta}] \\ &= \boldsymbol{xy}\otimes[\boldsymbol{\xi},\boldsymbol{\eta}] + [\boldsymbol{x},\boldsymbol{y}]\otimes\boldsymbol{\eta\xi} \end{aligned} \tag{93}$$

In several ways, we can make $CLW(V)$ into a superalgebra. But, for compatibility of grading and the natural inner products on $CLW(V)$ that is induced by the inner products on $CL(V_0)$ and $WL(V_1)$, we choose the following grading for $CLW(V)$.

$$\varepsilon(\boldsymbol{x}\otimes\boldsymbol{\xi}\ ) = \varepsilon(\boldsymbol{\xi}\ ) \tag{94}$$

In this grading, even and odd subspaces of $CLW(V)$ are as follows.

$$CLW(V)_0 = CL(V_0)\otimes WL(V_1)_0 \quad , \quad CLW(V)_1 = CL(V_0)\otimes WL(V_1)_1 \tag{95}$$

For this grading, the super Lie bracket of simple elements of $CLW(V)$ satisfies the following relation.

$$\begin{aligned} [\boldsymbol{x}\otimes\boldsymbol{\xi}\ ,\boldsymbol{y}\otimes\boldsymbol{\eta}\ ]^s &= (\boldsymbol{x}\otimes\boldsymbol{\xi})(\boldsymbol{y}\otimes\boldsymbol{\eta}) - (-1)^{\xi\eta}(\boldsymbol{y}\otimes\boldsymbol{\eta})(\boldsymbol{x}\otimes\boldsymbol{\xi}) \\ &= (-1)^{\xi y}\boldsymbol{xy}\otimes\boldsymbol{\xi\eta} - (-1)^{\xi\eta + x\eta}\boldsymbol{yx}\otimes\boldsymbol{\eta\xi} \end{aligned} \tag{96}$$

In some special cases, this cumbersome relation can be written simpler. If both $\boldsymbol{x}, \boldsymbol{\xi}$ are even, then:

$$[\boldsymbol{x} \otimes \boldsymbol{\xi}\ , \boldsymbol{y} \otimes \boldsymbol{\eta}\ ]^s = \boldsymbol{yx} \otimes [\boldsymbol{\xi}, \boldsymbol{\eta}] - [\boldsymbol{y}, \boldsymbol{x}] \otimes \boldsymbol{\xi\eta} \tag{97}$$

If both $\boldsymbol{x}, \boldsymbol{\xi}$ are odd, then:

$$[\boldsymbol{x} \otimes \boldsymbol{\xi}\ , \boldsymbol{y} \otimes \boldsymbol{\eta}\ ]^s = \boldsymbol{yx} \otimes [\boldsymbol{\xi}, \boldsymbol{\eta}] - [\boldsymbol{y}, \boldsymbol{x}]^s \otimes \boldsymbol{\xi\eta} \tag{98}$$

The restrictions of this super Lie bracket on $CL(V_0)$ and $WL(V_1)$ are respectively the Lie bracket of $CL(V_0)$ and the super Lie bracket of $L(V_1)$ .

The inner products on $CL(V_0)$ and $WL(V_1)$, naturally induce the following inner product on $CLW(V)$ that is super symmetric for the above grading of $CLW(V)$.

$$< \boldsymbol{x} \otimes \boldsymbol{\xi}\ , \boldsymbol{y} \otimes \boldsymbol{\eta}\ >= (-1)^{\xi y} < \boldsymbol{x}\ , \boldsymbol{y}\ >_{\text{new}} < \boldsymbol{\xi}\ , \boldsymbol{\eta}\ > \tag{99}$$

In other words, this inner product is symmetric on $CLW(V)_0$ and is alternating on $CLW(V)_1$ and these subspaces are orthogonal to each other.

**Theorem** 16. For arbitrary elements of $CLW(V)$ such as $\boldsymbol{X}, \boldsymbol{Y}, \boldsymbol{Z}$, we have:

$$< \boldsymbol{XY}, \boldsymbol{Z} >= (-1)^{X} < \boldsymbol{Y}, \boldsymbol{ZX} > \ , \ < \boldsymbol{XY}, \boldsymbol{Z} >=< \boldsymbol{X}, \boldsymbol{YZ} > \tag{100}$$

**Proof**. We can assume $\varepsilon(\boldsymbol{X}) + \varepsilon(\boldsymbol{Y}) + \varepsilon(\boldsymbol{Z}) = 0$, because in other cases, both sides of (100) is zero. Assume $\boldsymbol{X} = \boldsymbol{x} \otimes \boldsymbol{\xi}\ , \boldsymbol{Y} = \boldsymbol{y} \otimes \boldsymbol{\eta}\ , \boldsymbol{Z} = \boldsymbol{z} \otimes \boldsymbol{\phi}$ .

$$\begin{aligned} < \boldsymbol{XY}, \boldsymbol{Z} > &=< (\boldsymbol{x} \otimes \boldsymbol{\xi}\ )(\boldsymbol{y} \otimes \boldsymbol{\eta}), \boldsymbol{z} \otimes \boldsymbol{\phi} >=< (-1)^{\xi y} \boldsymbol{xy} \otimes \boldsymbol{\xi\eta}\ , \boldsymbol{z} \otimes \boldsymbol{\phi} > \\ &= (-1)^{\xi y + (\xi+\eta) z} < \boldsymbol{xy}\ , \mathbf{z}\ >_{\text{new}} < \boldsymbol{\xi}\ \boldsymbol{\eta}\ , \boldsymbol{\phi}\ > \\ &= (-1)^{\xi y + (\xi+\eta) z} < \boldsymbol{x}, \boldsymbol{y}\mathbf{z}\ >_{\text{new}} < \boldsymbol{\xi}\ , \boldsymbol{\eta}\ \boldsymbol{\phi}\ > \\ &= (-1)^{\xi y + (\xi+\eta) z + \eta z + \xi(y+z)} < (\boldsymbol{x} \otimes \boldsymbol{\xi}\ ), (\boldsymbol{y} \otimes \boldsymbol{\eta})(\boldsymbol{z} \otimes \boldsymbol{\phi}) > \\ &=< (\boldsymbol{x} \otimes \boldsymbol{\xi}\ ), (\boldsymbol{y} \otimes \boldsymbol{\eta})(\boldsymbol{z} \otimes \boldsymbol{\phi}) > =< \boldsymbol{X}, \boldsymbol{YZ} > \end{aligned} \tag{101}$$

$$\begin{aligned} < \boldsymbol{XY}, \boldsymbol{Z} > &= (-1)^{\xi y + (\xi+\eta) z} < \boldsymbol{xy}\ , \mathbf{z}\ >_{\text{new}} < \boldsymbol{\xi}\ \boldsymbol{\eta}\ , \boldsymbol{\phi}\ > \\ &= (-1)^{\xi y + (\xi+\eta) z + \xi} < \boldsymbol{y}, \mathbf{z}\boldsymbol{x} >_{\text{new}} < \boldsymbol{\eta}\ , \boldsymbol{\phi\xi}\ > \\ &= (-1)^{\xi y + (\xi+\eta) z + \xi + x\phi + \eta(x+z)} < (\boldsymbol{y} \otimes \boldsymbol{\eta}), (\boldsymbol{z} \otimes \boldsymbol{\phi})(\boldsymbol{x} \otimes \boldsymbol{\xi}\ ) > \\ &= (-1)^{\xi} < (\boldsymbol{y} \otimes \boldsymbol{\eta}), (\boldsymbol{z} \otimes \boldsymbol{\phi})(\boldsymbol{x} \otimes \boldsymbol{\xi}\ ) > = (-1)^{X} < \boldsymbol{Y}, \boldsymbol{ZX} > \ \blacksquare \end{aligned}$$

**Cororally** 17. For arbitrary elements of $CLW(V)$ such as $\boldsymbol{X}, \boldsymbol{Y}, \boldsymbol{Z}$ , we have:

$$< [\boldsymbol{X},\boldsymbol{Y}]^s, \boldsymbol{Z} >=< \boldsymbol{X}, [\boldsymbol{Y},\boldsymbol{Z}]^s > \tag{102}$$

For a simple element of $CLW(V)$ like $\boldsymbol{x} \otimes \boldsymbol{\xi}$, its order is defined by the sum of the order of $\boldsymbol{x}$ and $\boldsymbol{\xi}$. The subspace of elements of order $k$ in $CLW(V)$ is denoted by $CLW^{(k)}(V)$. So,

$$CLW^{(k)}(V) = \oplus_{r+s=k} \Lambda^r V_0 \otimes \vee^s V_1 \tag{103}$$

For example:

$$\begin{aligned} CLW^{(1)}(V) &= V_0 \oplus V_1 = V \\ CLW^{(2)}(V) &= \Lambda^2 V_0 \oplus \vee^2 V_1 \oplus (V_0 \otimes V_1) \end{aligned} \tag{104}$$

For any $\boldsymbol{X} \in CLW(V)$ the operator $\mathrm{ad}(\boldsymbol{X}): CLW(V) \longrightarrow CLW(V)$ is defined by $\mathrm{ad}(\boldsymbol{X})(\boldsymbol{Y}) = [\boldsymbol{X},\boldsymbol{Y}]^s$. If the order of $\boldsymbol{X}$ is 2, then all subspaces $CLW^{(k)}(V)$ are invariant under $\mathrm{ad}(\boldsymbol{X})$. Especially, $CLW^{(2)}(V)$ is a sub Lie superalgebra, and the action of $CLW^{(2)}(V)$ on $CLW^{(k)}(V)$ is a representation of this super Lie algebra.

An important point here is that the Lie superalgebra $CLW^{(2)}(V)$ is naturally isomorphic to $\mathfrak{osp}(V)$ (ortho-symplectic Lie superalgebra) via the following isomorphism.

$$\begin{aligned} CLW^{(2)}(V) &\longrightarrow \mathfrak{osp}(V) \\ \mathbf{X} &\longmapsto \mathrm{ad}(\boldsymbol{X})|_V \end{aligned} \tag{105}$$

Via this isomorphism, we find that:

$$\mathfrak{osp}_0(V) = \Lambda^2(V_0) \oplus \vee^2(V_1) \ , \ \mathfrak{osp}_1(V) = V_0 \otimes V_1 \tag{106}$$

**Cororally** 18. For any $T \in \mathfrak{osp}(V)$ there exists a unique $\boldsymbol{X} \in CLW^{(2)}(V)$ such that:

$$\forall z \in V \quad T(z) = [\boldsymbol{X}, z]^s = [\boldsymbol{X}, z] \tag{107}$$

## 5. **Star structures on the Clifford-Weyl algebras**

The star structures are defined only in complex spaces. In this section, all spaces are complex. For physical applications in antiparticles representation [3], we need to introduce star structures into the scene.

For a (complex) space $V$, a mapping $x \mapsto x^*$on $V$ is called a star structure on $V$ if it is conjugate linear and is its inverse, i.e. $x^{**} = x$. For example, for any real space $W$, $W^{\mathbb{C}}$ has a natural star structure $u + iv \mapsto u - iv$. It can be seen that every star structure comes from this example. In fact, for a complex space $V$ which has a star structure, the set of selfadjoint vectors (i.e. $x^* = x$) form a real subspace $W$, which $W^{\mathbb{C}}$ is naturally isomorphic to $V$, and under this isomorphism, the star operator on $V$ transforms to the natural star operator on $W^{\mathbb{C}}$

A complex superspace with a star structure on its underlying space is called a star superspace whenever $\varepsilon(x^*) = \varepsilon(x)$. In this case, we can restrict the star operator to the even and odd subspaces that make them to have a star structure. On the contrary, having separate star structures on the even and odd subspaces is equivalent to have a star structure on the whole superspace.

A star algebra is an algebra with a star structure on its underlying vector space such that $(xy)^* = y^*x^*$. But, a star superalgebra is a superalgebra with a star structure on its underlying supervector space such that $(xy)^* = (-1)^{xy}y^*x^*$

If V has a star structure and is equipped with a symmetric or alternating inner product, we say the star structure is compatible with the inner product, whenever:

$$< x^*, y^* >=\overline{< x, y >} \tag{108}$$

Any star vector space that has a compatible symmetric or alternating inner product, is made by complexification of a real space with symmetric or alternating real inner product.

In the case of symmetric inner products that have a compatible star structure, the following operation creates a Hermitian inner product in the space.

$$(x, y) \mapsto (x|y) =< x, y^* > \tag{109}$$

This Hermitian inner product are not necessarily positive definite. The relationship between the star operator and this Hermitian inner product is $(x^*|y^*) = \overline{(x|y)}$.

But in the case of alternating inner products that have a compatible star structure, the following operation creates a Hermitian multiplication on the space.

$$(x, y) \mapsto (x|y) = i < x, y^* > \tag{110}$$

In this case, this Hermitian inner product is necessarily indefinite and its relation with the star operator is $(x^*|y^*) = -\overline{(x|y)}$.

In superspaces that have a star structure and a supersymmetric inner product, we say that the star operator is compatible with the inner product whenever:

$$< x^*, y^* > = \overline{< x, y >} \tag{111}$$

So, we can define a natural Hermitian inner products on the superspace as follows.

$$(x|y) = \begin{cases} < x, y^* > & \text{one of } x, y \text{ is even} \\ i < x, y^* > & \text{both } x, y \text{ are odd} \end{cases} \tag{112}$$

Even and odd subspaces are orthogonal to each other and the star operator has the following properties for this Hermitian inner product.

$$(x^*|y^*) = (-1)^{xy}\overline{(x|y)} \tag{113}$$

If a vector space $V$ has a star structure, a star structure is induced on the exterior algebra $\wedge V$ as follows.

$$(x_1 \wedge \cdots \wedge x_k)^* = x_k^* \wedge \cdots \wedge x_1^* \tag{114}$$

So, we have; $(\boldsymbol{x} \wedge \boldsymbol{y})^* = \boldsymbol{y}^* \wedge \boldsymbol{x}^*$ . This star structure does not respect the superalgebra structure of $\wedge V$ but respects its ordinary algebra structure.

If a symmetric inner product on $V$ is compatible with its star structure, then the symmetric inner product that is induced on $\wedge V$ is compatible with the induced star structure. In this case, we can also construct a star operator on $CL(V)$ that make this algebra into a star algebra. To construct this star operator on $CL(V)$, consider the map $x \mapsto x^*$ from $V$ to $\overline{CL(V)}^{opp}$. This is a Clifford map and is extended to an algebra isomorphism from $CL(V)$ to $\overline{CL(V)}^{opp}$. This isomorphism as a map on $CL(V)$ is denoted by $\boldsymbol{x} \mapsto \boldsymbol{x}^*$ and makes $CL(V)$ into a star algebra.

The natural isomorphism between $\wedge V$ and $CL(V)$ respects the star structures on these algebras, so these star structures are the same. Therefore, the symmetric inner products on $CL(V)$ are also compatible with its star structure.

The relation between the star operator and the Lie bracket on $CL(V)$ is as follows.

$$[\boldsymbol{x}, \boldsymbol{y}]^* = -[\boldsymbol{x}^*, \boldsymbol{y}^*] \tag{115}$$

If $V$ has a star structure, then $\vee V$ as a superalgebra will have a star structure as follows.

$$(\xi_1 \vee \cdots \vee \xi_k)^* = (-1)^{\frac{k(k-1)}{2}} \xi_1^* \vee \cdots \vee \xi_k^* \tag{116}$$

This definition implies the following relation.

$$(\boldsymbol{\xi} \vee \boldsymbol{\eta})^* = (-1)^{\xi\eta} \boldsymbol{\eta}^* \vee \boldsymbol{\xi}^* \tag{117}$$

If an alternating inner product on $V$ is compatible with its star structure, then the super symmetric inner product that is induced on the superalgebra $\vee V$ is compatible with the induced star structure. In this case, we can also construct a star operator on $WL(V)$ that make this superalgebra into a star superalgebra.

To construct this star operator on $WL(V)$, first, we remind that for each superalgebra $A = A_0 \oplus A_1$, the notation $A^{\mathrm{sopp}}$ refers to the superspace $A$ which is considered by the following new multiplication: $a.b = (-1)^{ab} ba$ By this multiplication, $A^{\mathrm{sopp}}$ is also a superalgebra. Now, consider the map $x \mapsto x^*$ from $V$ to $\overline{WL(V)^{sopp}}$. This is a Weyl map and is extended to an algebra isomorphism from $WL(V)$ to $\overline{WL(V)^{sopp}}$. This isomorphism as a map on $WL(V)$ is denoted by $\boldsymbol{\xi} \mapsto \boldsymbol{\xi}^*$ and makes $WL(V)$ into a star superalgebra.

The natural isomorphism between $\vee V$ and $WL(V)$ respects the star structures on these superalgebras, so these star structures are the same. Therefore, the super-symmetric inner products on $WL(V)$ are also compatible with its star structure.

The relation between the star operator and the super Lie bracket on $WL(V)$ is as follows.

$$[\boldsymbol{\xi}, \boldsymbol{\eta}]^{s*} = -[\boldsymbol{\xi}^*, \boldsymbol{\eta}^*]^s \tag{118}$$

If $V = V_0 \oplus V_1$ is a superspace with a supersymmetric inner product and a compatible star structure, we can extend its star operator on $CLW(V)$ such that it becomes a star superalgebra. To this end, consider the mapping $x \longmapsto x^*$ from $V$ to $\overline{CLW(V)^{sopp}}$ which is a Clifford-Weyl map and extends to an algebra isomorphism from $CLW(V)$ to $\overline{CLW(V)^{sopp}}$. This extended mapping is denoted by the same symbol *. Restrictions of this star operator on $CL(V_0)$ and $WL(V_1)$ are the same as the star operators defined above and in general:

$$(\boldsymbol{x} \otimes \boldsymbol{\xi})^* = (-1)^{x\xi} \boldsymbol{x}^* \otimes \boldsymbol{\xi}^* \tag{119}$$

As we saw in (112), we can define a Hermitian inner product on $CLW(V)$ such that

$$(\boldsymbol{X}^*|\boldsymbol{Y}^*) = (-1)^{XY}\overline{(\boldsymbol{X}|\boldsymbol{Y})} \tag{120}$$

This Hermitian inner product for simple elements is done as follows.

$$(\boldsymbol{x} \otimes \boldsymbol{\xi}|\boldsymbol{y} \otimes \boldsymbol{\eta}) = (\boldsymbol{x}|\boldsymbol{y})(\boldsymbol{\xi}|\boldsymbol{\eta}) \tag{121}$$

## 6. **Roots and weights for ortho-symplectic Lie superalgebras**

**Orthogonal Lie algebras**: For a complex space $V$ and a symmetric inner product on it, the Lie algebra $\mathfrak{o}(V)$ has a natural imbedding in $CL(V)$ and is isomorphic to $\Lambda^2 V$. By using the Lie algebra structure of $CL(V)$, we can naturally find roots of $\mathfrak{o}(V)$ and weights of representations of $\mathfrak{o}(V)$ on $\Lambda^k V$ and the whole of $CL(V)$. We can assume $V$ has a compatible star structure and the restriction of the inner product on the real subspace of selfadjoint vectors is positive definite. So, there exists an orthonormal base of selfadjoint vectors like $\{v_1, \cdots, v_d\}$ (i.e. $v_i^* = v_i, < v_i, v_j >= \delta_{ij}$). First, suppose that $d$ is even and $d = 2n$. In this case, we can construct a base in the form $\{e_1, \cdots, e_n, e_1^*, \cdots, e_n^*\}$ such that:

$$< e_i, e_j >=< e_i^*, e_j^* >= 0 \quad , \quad < e_i, e_j^* >= \delta_{ij} \tag{122}$$

For example, put :

$$e_1 = \frac{1}{\sqrt{2}}(v_1 + iv_2) \, , \; \dots \; , \; e_n = \frac{1}{\sqrt{2}}(v_{2n-1} + iv_{2n}) \tag{123}$$

Put $H_i = \frac{1}{2} e_i \wedge e_i^*, i = 1, \cdots, n$ and $\mathfrak{h} = \mathrm{span}\{H_1, \cdots, H_n\}$. Since $\{H_1, \cdots, H_n\}$ is linearly independent and $[H_i, H_j] = 0$, $\mathfrak{h}$ is a commutative Lie subalgebra of $\Lambda^2 V$ and is $n$-dimensional, therefore, it is a Cartan subalgebra of $\Lambda^2 V$ [7]. The dual base of $\{H_1, \cdots, H_n\}$ is denoted by $\{H^1, \cdots, H^n\}$. For the representation of $\wedge^2 V$ on $V$, we find that:

$$\begin{aligned} \mathrm{ad}(H_t)(e_i) &= [H_t, e_i] = \delta_{it} e_i = H^i(H_t) e_i \\ \mathrm{ad}(H_t)(e_i^*) &= [H_t, e_i^*] = -\delta_{it} e_i^* = -H^i(H_t) e_i^* \end{aligned} \tag{124}$$

These equations mean that $\pm H^i$ are the weights of this representation and the eigen subspaces of these weights are also one-dimensional [7]. From these relations, we can obtain important results about the weights of the representation of $\wedge^2 V$ on $\wedge^m V$.

Note that from (45) for $\boldsymbol{x} \in \wedge^2 V$ we have:

$$\mathrm{ad}(\boldsymbol{x})(y_1 \wedge \cdots \wedge y_m) = \sum_{i=1}^{m} y_1 \wedge \cdots \wedge \mathrm{ad}(\boldsymbol{x})(y_i) \wedge \cdots \wedge y_m \tag{125}$$

For simplicity for any vector $x \in V$ and $\alpha = 0,1$ , put:

$$x^\alpha = \begin{cases} x & \alpha = 1 \\ 1 & \alpha = 0 \end{cases} \tag{126}$$

For any $k_1, \cdots, k_n, l_1, \cdots, l_n \in \{0,1\}$ such that $k_1 + \cdots + k_n + l_1 + \cdots + l_n = m$ the following combination of $H^i$,s is a weight of the representation of $\wedge^2 V$ on $\wedge^k V$.

$$(k_1 - l_1)H^1 + \cdots + (k_n - l_n)H^n \tag{127}$$

The following $k$-multivector is an eigenvector for this eigenfunctional.

$$e_1^{k_1} \wedge \cdots \wedge e_n^{k_n} \wedge (e_1^*)^{l_1} \wedge \cdots \wedge (e_n^*)^{l_n} \tag{128}$$

Especially, in the case $k = 2$, nonzero weights are roots of $\wedge^2 V$. So, the roots of $\wedge^2 V$ are the following functionals.

$$\pm(H^i + H^j) \ , \ \pm(H^i - H^j) \qquad (1 \leq i < j \leq n) \tag{129}$$

Now, suppose $V$ is odd-dimensional and $d = 2n + 1$. In this case, by using an orthonormal base like $\{v_1, \cdots, v_{2n}, e\}$, in the same way, we can construct a base $\{e_1, \cdots, e_n, e_1^*, \cdots, e_n^*, e\}$ such that (122) is satisfied and $< e_i, e >=< e_i^*, e >= 0$.

We use the same definitions for $H_i$,s and put $\mathfrak{b} = \mathrm{span}\{H_1, \cdots, H_n\}$. So, $\mathfrak{b}$ is a Cartan subalgebra of $\wedge^2 V$ too. The following relation hold.

$$\begin{aligned} \mathrm{ad}(H_t)(e_i) &= H^i(H_t)e_i \\ \mathrm{ad}(H_t)(e_i^*) &= -H^i(H_t)e_i^* \\ \mathrm{ad}(H_t)(e) &= 0 \end{aligned} \tag{130}$$

These equations mean that $\pm H^i$ and zero functional are the weights of this representation and the eigensubspaces of these weights are also one-dimensional. From these relations, we can also find the weights of the representation of $\wedge^2 V$ on $\wedge^m V$.

For any $k_1, \cdots, k_n, l_1, \cdots, l_n, l \in \{0,1\}$ such that $k_1 + \cdots + k_n + l_1 + \cdots + l_n + l = m$ the following combination of $H^i$,s is a weight of the representation of $\wedge^2 V$ on $\wedge^m V$.

$$(k_1 - l_1)H^1 + \cdots + (k_n - l_n)H^{\mathrm{n}} + \mathrm{l} \times 0 \tag{131}$$

The following $k$-multivector is an eigenvector for this eigenfunctional.

$$e_1^{k_1} \wedge \cdots \wedge e_n^{k_n} \wedge (e_1^*)^{l_1} \wedge \cdots \wedge (e_n^*)^{l_n} \wedge \mathrm{e}^l \tag{132}$$

Especially, in the case $k = 2$, roots of $\wedge^2 V$ are nonzero weights, so roots of $\wedge^2 V$ are the following functionals.

$$\pm(H^i + H^j)\ ,\ \pm(H^i - H^j),\ \pm H^i \qquad (1 \leq i < j \leq n) \tag{133}$$

**Symplectic Lie algebras**: For a complex space $V$ and an alternating inner product on it, the Lie algebra $\mathfrak{sp}(V)$ has a natural imbedding in $WL(V)$ and is isomorphic to $\vee^2 V$. By using the Lie algebra structure of $WL(V)$, we can naturally find roots of $\mathfrak{sp}(V)$ and weights of representations of $\mathfrak{sp}(V)$ on $\vee^m V$ and the whole of $\mathrm{WL}(V)$. Suppose $\{\xi_1, \cdots, \xi_n, \xi^1, \cdots, \xi^n\}$ is a symplectic base for $V$, that means:

$$< \xi_i, \xi_j >=< \xi^i, \xi^j >= 0\ , < \xi_i, \xi^j >= \delta_i^j \tag{134}$$

Put $K_i = \frac{1}{2}\xi_i \vee \xi^i, i = 1, \cdots, n$ . By (81) we find that $[K_i,\ K_j] = 0$ and the subspace $\mathfrak{b} = \mathrm{span}\{K_1, \cdots, K_n\}$ is a Cartan subalgebra of $\vee^2 V$. Dual base of $\{K_1, \cdots, K_n\}$ is denoted by $\{K^1, \cdots, K^n\}$ and for the representation of $\vee^2 V$ on $V$, we find that:

$$\begin{aligned} \mathrm{ad}(K_t)(\xi_i) &= [K_t, \xi_i] = -\delta_{it}\xi_i = -K^i(K_t)\xi_i \\ \mathrm{ad}(K_t)(\xi^i) &= [K_t, \xi^i] = \delta_{it}\xi^i = K^i(K_t)\xi^i \end{aligned} \tag{135}$$

These equations mean that $\pm K^i$are the weights of this representation and the eigen subspaces of these weights are also one-dimensional. From these relations, we can compute the weights of the representation of $\vee^2 V$ on $\vee^m V$.

Note that from (81) we can deduce the following relation for every $\boldsymbol{\xi} \in \vee^2 V$.

$$ad(\boldsymbol{\xi})(\eta_1 \vee \cdots \vee \eta_k) = \sum_{i=1}^{k} \eta_1 \vee \cdots \vee ad(\boldsymbol{\xi})(\eta_i) \vee \cdots \vee \eta_k \tag{136}$$

For each $\xi \in V$, put $\xi^k = \xi \vee \cdots \vee \xi$ ( $k$ term). For any non-negative integer $k_1, \cdots, k_n, l_1, \cdots, l_n$ such that $k_1 + \cdots + k_n + l_1 + \cdots + l_n = m$ the following combination of $K^i$,s is a weight of the representation of $\vee^2 V$ on $\vee^m V$.

$$(k_1 - l_1)K^1 + \cdots + (k_n - l_n)K^n \tag{137}$$

Because, the symmetric multivector $\xi_1^{l_1} \vee \cdots \vee \xi_n^{l_n} \vee (\xi^1)^{k_1} \vee \cdots \vee (\xi^n)^{k_n}$ whose order is $m$, is an eigenvector for this functional. So, for any integer $a_1, \cdots, a_n$ such that $|a_1| + \cdots + |a_n| \leq m$ and $|a_1| + \cdots + |a_n| - m$ is even, the functional $a_1 K^1 + \cdots + a_n K^n$ is a weight for this representation. In particular, the roots of $\vee^2 V$ are the following functional.

$$\pm(K^i + K^j),\ \pm(K^i - K^j)\ , \pm 2K^i\ \ (1 \leq i < j \leq m) \tag{138}$$

Moreover, for the representation of $\vee^2 V$ on $WL(V)$ any integer combination of $K^i$,s is a weight and the eigensubspace of any weight is infinite dimensional.

**Orthosymplectic Lie superalgebras**: For a complex superspace $V = V_0 \oplus V_1$ and a supersymmetric inner product on it, the Lie superalgebra $\mathfrak{osp}(V)$ [2] has a natural imbedding in $CLW(V)$ and is isomorphic to $CLW^{(2)}V$. By using the Lie superalgebra structure of $CLW(V)$, we can naturally find roots of $\mathfrak{osp}(V)$ and weights of representations of $\mathfrak{osp}(V) = \wedge^2 V_0 \oplus \vee^2 V_1 \oplus (V_0 \otimes V_1)$ on $CLW^{(k)}V$ and whole of $CLW(V)$.

First suppose $V_0$ is even dimensionaland $\dim(V_0) = 2n, \dim(V_1) = 2m$. For a suitable star structure, we can consider a base $\{e_1, \cdots, e_n, e_1^*, \cdots, e_n^*\}$ for $V_0$ and a base $\{\xi_1, \cdots, \xi_n, \xi^1, \cdots, \xi^n\}$ for $V_1$ such that:

$$< e_i, e_j >=< e_i^*, e_j^* >= 0 \ ,< e_i, e_j^* >= \delta_{ij}$$
$$< \xi_i, \xi_j >=< \xi^i, \xi^j >= 0 \ ,< \xi_i, \xi^j >= \delta_i^j \tag{139}$$

Put $H_i = \frac{1}{2} e_i \wedge e_i^*, (i = 1, \cdots, n)$ , $K_j = \frac{1}{2} \xi_j \vee \xi^j, (j = 1, \cdots, m)$. These are some vectors in $\mathfrak{osp}_0(V)$ and $\mathfrak{b} = \text{span}\{H_1, \cdots, H_n, K_1, \cdots, K_m\}$ is a Cartan subalgebra for $\mathfrak{sp}_0(V)$. If $\mathfrak{b}_1 = \text{span}\{H_1, \cdots, H_n\}$ and $\mathfrak{b}_2 = \text{span}\{K_1, \cdots, K_m\}$, then $\mathfrak{b} = \mathfrak{b}_1 \oplus \mathfrak{b}_2$ and $\mathfrak{b}^* = \mathfrak{b}_1^* \oplus \mathfrak{b}_2^*$ .

So, the functionals $H^1, \cdots, H^n, K^1, \cdots, K^m$ , that are defined above, can be considered as functionals on $\mathfrak{b}$ and is the dual base of $\{H_1, \cdots, H_n, K_1, \cdots, K_m\}$. Moreover, for $\alpha \in \mathfrak{b}_1^*$ and $\beta \in \mathfrak{b}_2^*$ and $u \in \mathfrak{b}_1$ and $v \in \mathfrak{b}_2$ we have:

$$(\alpha + \beta)(u + v) = \alpha(u) + \beta(v) \tag{140}$$

All subspaces $\wedge^r V_0 \otimes \vee^s V_1$ of $CLW(V)$ are invariant under $\text{ad}(\boldsymbol{X})$ for $\boldsymbol{X} \in \mathfrak{osp}_0(V)$, so they are a representation for $\mathfrak{osp}_0(V)$. By (97) we can find the relation between the representation of $\wedge^2 V_0$ on $\wedge^r V_0$ and the representation of $\vee^2 V_1$ on $\vee^s V_1$ and the representation of $\mathfrak{osp}_0(V)$ on $\wedge^r V_0 \otimes \vee^s V_1$ as follows. For $\boldsymbol{x} \in \wedge^2 V_0$ and $\boldsymbol{\xi} \in \vee^2 V_1$ and $\boldsymbol{y} \in \wedge^r V_0$ and $\boldsymbol{\eta} \in \vee^s V_1$ we have:

$$ad(\boldsymbol{x})(\boldsymbol{y} \otimes \boldsymbol{\eta}) = ad(\boldsymbol{x})(\boldsymbol{y}) \otimes \boldsymbol{\eta} \ , \ ad(\boldsymbol{\xi})(\boldsymbol{y} \otimes \boldsymbol{\eta}) = \boldsymbol{y} \otimes ad(\boldsymbol{\xi})(\boldsymbol{\eta}) \tag{141}$$

This relation shows that if $\alpha$ is a weight for the representation of $\wedge^2 V_0$ on $\wedge^r V_0$ and $\beta$ is a weight for the representation of $\vee^2 V_1$ on $\vee^s V_1$, then $\alpha + \beta$ is a weight for the representation of $\mathfrak{osp}_0(V)$ on $\wedge^r V_0 \otimes \vee^s V_1$. In fact, if $\boldsymbol{y} \in \wedge^r V_0$ is an eigenvector for $\alpha$ and $\boldsymbol{\eta} \in \vee^s V_1$ is an eigenvector for $\beta$ then $\boldsymbol{y} \otimes \boldsymbol{\eta}$ is an eigenvector for $\alpha + \beta$. Because for any $h = h_1 + h_2 \in \mathfrak{b} = \mathfrak{b}_1 \oplus \mathfrak{b}_2$ we have:

$$\begin{aligned} ad(h)(\boldsymbol{y} \otimes \boldsymbol{\eta}) &= ad(h_1)(\boldsymbol{y} \otimes \boldsymbol{\eta}) + ad(h_2)(\boldsymbol{y} \otimes \boldsymbol{\eta}) \\ &= ad(h_1)(\boldsymbol{y}) \otimes \boldsymbol{\eta} + \boldsymbol{y} \otimes ad(h_2)(\boldsymbol{\eta}) \\ &= \alpha(h_1) \boldsymbol{y} \otimes \boldsymbol{\eta} + \beta(h_2) \boldsymbol{y} \otimes \boldsymbol{\eta} = (\alpha + \beta)(h)(\boldsymbol{y} \otimes \boldsymbol{\eta}) \end{aligned} \tag{142}$$

Since $\mathfrak{osp}_0(V) = \wedge^2 V_0 \oplus \vee^2 V_1$, the set of even roots of $\mathfrak{osp}(V)$ (i.e. the set of roots of $\mathfrak{osp}_0(V)$) is the union of the sets of roots of $\wedge^2 V_0$ and $\vee^2 V_1$.

$$\begin{array}{ll} \pm(H^i + H^j)\ ,\ \pm(H^i - H^j) & (1 \le i < j \le n) \\ \pm(K^i + K^j),\ \pm(K^i - K^j)\ ,\pm 2K^{\mathrm{i}} & (1 \le i < j \le m) \end{array} \tag{143}$$

Also, odd roots of $\mathfrak{osp}(V)$ are weights of adjoint representation of $\mathfrak{osp}_0(V)$ on $\mathfrak{osp}_1(V) = V_0 \otimes V_1$. So, by (142) we find the odd roots of $\mathfrak{osp}(V)$ by adding weights of representation $\wedge^2 V_0$ on $V_0$ and wights of representation of $\vee^2 V_1$ on $V_1$ to each other.

$$\pm H^i \pm K^j \quad i = 1, \cdots, n\ \ j = 1, \cdots, m \tag{144}$$

In this case, all odd roots are isotropic. In general, an odd root $\alpha$ is nonisotropic iff $2\alpha$ is an even root. We can find even and odd weights of representation $\mathfrak{osp}(V)$ on $CLW^{(m)}(V)$ in the same way.

Now, suppose $V_0$ is odd dimensional and $\dim(V_0) = 2n + 1$. Using the same notation as before, consider a base $\{e_1, \cdots, e_n, e_1^*, \cdots, e_n^*, e\}$ for $V_0$ such that:

$$< e_i, e_j > = < e_i^*, e_j^* > = 0\ , < e_i, e_j^* > = \delta_{ij}, < e_i, e > = < e_i^*, e > = 0$$

As before, we construct $H^i$,s and the roots of $\wedge^2 V$ are the following functionals.

$$\pm(H^i + H^j)\ ,\ \pm(H^i - H^j),\ \ \pm H^i \qquad (1 \le i < j \le n)$$

Now, the set of even the roots of $\mathfrak{osp}(V)$ is the union of the sets of roots of $\wedge^2 V_0$ and $\vee^2 V_1$.

$$\begin{array}{ll} \pm(H^i + H^j)\ ,\ \pm(H^i - H^j)\ \ \pm H^i & (1 \le i < j \le n) \\ \pm(K^i + K^j),\ \pm(K^i - K^j)\ ,\pm 2K^{\mathrm{i}} & (1 \le i < j \le m) \end{array} \tag{145}$$

By (142) the odd roots of $\mathfrak{osp}(V)$ are as follows.

$$\pm H^i \pm K^j\ ,\pm K^j\ \ i = 1, \cdots, n\ \ j = 1, \cdots, m \tag{146}$$

Here, $\pm K^j$,s are non-isotropic roots, and the rest are isotropic.

## References

[1] Carmeli Claudio, Lauren Caston, Rita Fioresi, *Mathematical Foundations of Supersymmetry*, European Mathematical Society, 2011

[2] Farmer Richard Joseph, *Orthosymplectic superalgebras in mathematics and science*, PhD Thesis (1984) https://figshare.utas.edu.au/ndownloader/files/40947935

[3 ] Gording Brage, *A Novel Approach To Particle Representations* arXiv:2005.06974v2 [physics.gen-ph] 7 Oct 2020

[4] Greub Werner. *Multiinear Algebra*, Springer-Verlag New York Inc (1978).ISBN-13:978-0-387-90284-5

[5] Hartwig, J.T., Serganova, V. *CLIFFORD AND WEYL SUPERALGEBRAS AND SPINOR REPRESENTATIONS*. Transformation Groups 25, 1185–1207 (2020). https://doi.org/10.1007/s00031-019-09542-7

[6] Kac V. G. *Lie Superalgebras* ADVANCES IN MATHEMATICS 26, 8-96 (1977)

[7] KIRILLOV ALEXANDER, Jr.*An Introduction to Lie Groups and Lie AlgebrasISBN-13 978-0-511-42319-2,* 2008

[8] Renaud Pierre, *CLIFFORD ALGEBRAS LECTURE NOTES ON APPLICATIONS IN PHYSICS,* https://www.mathematik.uni-muenchen.de › clifford

[9] Varadarajan Veeravalli , *Supersymmetry for mathematicians: An introduction*, Courant lecture notes in mathematics, American Mathematical Society Providence, R.I 2004 (doi:10.1090/cln/011)

[10] Woit, P. (2017). Weyl and Clifford Algebras. In: Quantum Theory, Groups and Representations. Springer, Cham. https://doi.org/10.1007/978-3-319-64612-1_28